\documentclass[preprint,showpacs,preprintnumbers,a4paper]{revtex4}
\usepackage{graphicx}% Include figure files
\usepackage{dcolumn}% Align table columns on decimal point
\usepackage{bm}% bold math
\usepackage{amssymb}
\usepackage{amsmath}
\usepackage{epsfig}    
\def\beq{\begin{equation}}
\def\eeq{\end{equation}}
\def\bea{\begin{array}}
\def\eea{\end{array}}
\def\be{\begin{equation}}
\def\ee{\end{equation}}
\def\ba{\begin{eqnarray}}
\def\ea{\end{eqnarray}}

\def\to{\rightarrow}

\def\[{\left[}
\def\]{\right]}
\def\({\left(}
\def\){\right)}

\def\sm0{{\widetilde{m}_0}}

\def\U1em{{U(1)_{\rm em}}}
\def\to{\rightarrow}

\def\sq2{\sqrt{2}}

\def\ee{e^+e^-}

\def\End{\end{document}}
\def\fsl#1{\setbox0=\hbox{$#1$}                 % set a box for #1 
   \dimen0=\wd0                                 % and get its size
   \setbox1=\hbox{/} \dimen1=\wd1               % get size of /
   \ifdim\dimen0>\dimen1                        % #1 is bigger
      \rlap{\hbox to \dimen0{\hfil/\hfil}}      % so center / in box
      #1                                        % and print #1
   \else                                        % / is bigger
      \rlap{\hbox to \dimen1{\hfil$#1$\hfil}}   % so center #1
      /                                         % and print /
   \fi}

\begin{document}                                                              
%\draft
%@twocolumnfalse\endcsname

%\title{Supersymmetric Higgs sectors in the decoupling limit}%
\title{Non-decoupling effects in supersymmetric Higgs sectors}%
\preprint{UT-HET 042, KU-PH-007}
\author{%
{\sc Shinya Kanemura\,$^1$, Tetsuo Shindou\,$^2$,  Kei Yagyu\,$^1$}
}
\affiliation{%
%\address{\vspace*{5mm}
\vspace*{2mm} 
$^1$Department of Physics, University of Toyama, 3190 Gofuku, Toyama 930-8555, Japan\\ 
$^2$Faculty of Engineering, 
    Kogakuin University, 
    1-24-2 Shinjuku, Tokyo 163-8677, Japan\\
}

\begin{abstract}

A wide class of Higgs sectors is 
investigated in supersymmetric standard models. %, 
When the lightest Higgs boson ($h$) looks the standard model one,
the mass ($m_h$) and the triple Higgs boson coupling (the $hhh$ coupling) 
are evaluated at the one-loop level in each model.
While $m_h$ is at most 120-130 GeV in the minimal 
supersymmetric standard model (MSSM), 
that in models with an additional neutral singlet or triplet fields can be much larger. 
The $hhh$ coupling can also be sensitive to the models:  
while  in the MSSM the deviation from the standard model prediction  
is not significant,  that can  be 
30-60 \%
in some models  
such as the MSSM with the additional singlet or with extra doublets 
and charged singlets.  
These models are motivated by specific physics problems like
the $\mu$-problem, the neutrino mass,
the scalar dark matter and so on.
Therefore, when $h$ is found at the CERN Large Hadron Collider, 
we can classify supersymmetric models 
by measuring $m_h$ and the $hhh$ coupling accurately 
at future collider experiments. 
\pacs{\, 14.80.Da, 12.60.-i, 12.60.Fr  }%\hfill~~[\today] }
\end{abstract}

\maketitle

\renewcommand{\thefootnote}{\arabic{footnote}}

\section{Introduction}

Physics of electroweak symmetry breaking (EWSB) 
is the last unknown part in the standard model (SM). 
Experimental identification of the Higgs sector has been one of the 
most important issues in high energy physics. 
Yet no Higgs boson has been found at the CERN LEP and 
the Fermilab Tevatron, 
the data are used to constrain the Higgs 
boson parameters~\cite{LEP,Tevatron}. 
The Higgs boson search has also started at the CERN Large Hadron Collider (LHC).
When a scalar boson is found in near future, its property will 
be measured as accurately as possible to see whether the scalar is really 
the Higgs boson. In particular, to understand the essence of EWSB, we must 
know the self-coupling in addition to the Higgs boson mass. 
The triple Higgs boson coupling may be explored at the LHC 
and its upgraded version\cite{hhh_lhc}, the International Linear 
Collider (ILC)\cite{hhh_lc} 
and its $\gamma\gamma$ option\cite{gammagamma} or the Compact
Linear Collider (CLIC).   

Experimental survey 
for the Higgs sector is important not only to confirm our picture 
for EWSB, but also to explore new physics beyond the SM (BSM). 
In the SM, appearance of the quadratic ultraviolet divergence 
in the one-loop calculation of the Higgs boson mass is 
a serious 
problem, which is called 
the hierarchy problem. 
This has to be eliminated in a model of BSM.
On the other hand, we already know several phenomena of BSM such as 
 dark matter, tiny neutrino masses, and baryon asymmetry of the universe. 

Supersymmetry (SUSY) is an attractive idea  
as a new physics scenario at the TeV scale. 
First of all, the hierarchy problem 
can be solved: i.e., the quadratic divergence 
due to a particle in the loop is 
cancelled by that due to the super partner particles. 
Second, a supersymmetric standard 
model (SSM) can naturally contain the candidate for 
dark matter; i.e., the lightest SUSY particle, 
whose stability may be guaranteed by the R-parity. 
One of the important 
predictions in the minimal SSM (MSSM), 
where the Higgs sector is composed of two Higgs doublets,  
is to the mass ($m_h$) of the lightest CP-even Higgs boson ($h$), which 
is predicted to be less than the $Z$ boson mass ($m_Z^{}$) at the tree level, but
can be maximally 30-40 GeV 
above $m_Z$ by radiative effects of the top quark and 
its super partners (stops)~\cite{mass_mssm}.
This prediction meets the present LEP bound on $m_h$~\cite{LEP}. 
If such a light $h$ will not be found, the MSSM must be ruled out. 
Is SUSY itself ruled out then? The answer is ``No''.
In fact, there can be several variations for the Higgs sector in SSMs.  
A simple extension may be addition of a neutral 
gauge singlet field to the MSSM, which is known as the Next-to  
MSSM (NMSSM)~\cite{nmssm,nmssm_mh,ellwanger}. 
Solving the $\mu$ problem~\cite{mu-problem} may be  
a motivation to the NMSSM.
It is known that $m_h$ in the NMSSM can be higher than that in the MSSM 
because of the additional contribution to $m_h$ from 
the tri-linear term of the Higgs doublets and the singlet in 
the superpotential. 
In addition to the MSSM and the NMSSM, a model with additional scalar boson
with higher representation can also push up $m_h$\cite{tripletmh}.
Furthermore, models with extended gauge sector can also have
higher $m_h$ than the MSSM prediction\cite{extragauge}, depending on the model parameters.
Therefore, $m_h$ can be an important tool to discriminate SSMs. 

In this Letter, we investigate Higgs sectors in a wide class of SSM,
when only the lightest Higgs boson $h$ 
appears at the electroweak (EW) scale and the other additional particles 
such as extra Higgs bosons and SUSY partner particles are rather heavier
(but not too heavy).
The coupling constants of $h$ to the SM particles are then similar  
to those in the SM at the tree level. 
In each model, $m_h$ as well as the deviation in the triple Higgs boson 
coupling (the $hhh$ coupling) are evaluated at the one-loop level. 
Their possible allowed values as well as their correlation are studied 
under the theoretical condition and the current experimental data. 
We here consider not only the MSSM and the NMSSM, but also 
further possible extensions of the MSSM. 
In order to widely examine various possibilities, we here   
do not necessarily require the paradigm of the grand unification. 
Instead, we may even retain the possibility 
of appearance of strong dynamics at the TeV scale as discussed 
in Ref.~\cite{fat_higgs}. % from the phenomenological interest. 

\section{Extended SUSY Models}
One way of the extension of the MSSM may be adding 
new chiral superfields  such as isospin singlets 
(neutral $S$, singly charged $\Omega_{\pm}$ or 
doubly charged $K_\pm$), doublets ($H_u'$ and $H_d'$), 
or triplets ($\xi$ with the hypercharge $Y=0$ or $\chi_\pm$ with $Y=\pm 1$), whose properties are defined
 in Table~\ref{field_prop}. 
As we are interested in the variation in the Higgs sector, 
these new fields are supposed to be colour singlet. 
Although there can be further possibilities such as introduction of 
new vector superfields which contain gauge fields 
for extra gauge symmetries, models 
with extra dimensions, those with the R-parity violation, etc, 
we here do not discuss them.
Eventually, in addition to the MSSM, we consider nine models  
in Table~\ref{tab:models}. 
%%%20110131
In the analysis of this letter, triple coupling terms with doublet Higgs 
superfields and additional chiral superfields play an important role.
In Table~\ref{tab:models}, such relevant terms are also listed.
%%%%%%
For anomaly cancellation, charged superfields are introduced in pair in each model. 

These extensions can be motivated by solving various physics problems. 
For example, models with additional charged singlet fields can be used 
for radiative neutrino mass generation~\cite{radiative_seesaw}.
Those with additional doublet fields may be required for dark doublet
models~\cite{dark_higgs}, 
and the model with triplets may be motivated for 
the SUSY left-right model\cite{LRmodels} or 
those with so-called the type-II seesaw 
mechanism~\cite{type-II_seesaw}.
Although models in Table~\ref{tab:models} can be imposed additional 
exact or softly-broken discrete symmetries for various  reasons,  
we here do not specify them as they do not affect our discussions.

%\section{The Higgs potential}

\begin{table}[t]
\begin{center}
  \begin{tabular}{|c|ccccc|cc|ccc|}
   \hline
& $S$      & $\Omega_+$ & $\Omega_-$ & $K_+$ & $K_-$ 
   & $H_u'$ & $H_d'$ & $\xi$ & $\chi_+$ & $\chi_-$  \\\hline
SU(2)$_I$ & 1 & 1& 1& 1& 1& 2& 2& 3& 3& 3  \\ \hline 
U(1)$_Y$ & 0 & 1& $-1$& 2& $-2$& $1/2$& $-1/2$& 0& 1& $-1$ \\ \hline 
  \end{tabular}
\end{center}
  \caption{Properties of additional chiral superfields.  }
  \label{field_prop}
 \end{table}

\begin{table}[t]
\begin{center}
  \begin{tabular}{|c|ccccc|cc|ccc||c|}
   \hline
%   & $H_u$  & $H_d$  
& $S$      & $\Omega_+$ & $\Omega_-$ & $K_+$ & $K_-$ 
   & $H_u'$ & $H_d'$ & $\xi$ & $\chi_+$ & $\chi_-$ &
   Relevant terms in the superpotential
   \\\hline
Model-1 & $\bullet$ & & & & & & & & &  
& $W\supset \lambda_{HHS}^{}H_u^{}\cdot H_d^{}S$\\ \hline 
Model-2 & & & & & & & &$\bullet$ & &
&$W\supset \lambda_{HH\xi}^{}H_u^{}\cdot \xi H_d^{}$\\ \hline 
Model-3 & &$\bullet$ &$\bullet$ & & & & & & &
&---\\ \hline 
Model-4 & &$\bullet$ &$\bullet$ &$\bullet$ &$\bullet$ & & & & & 
&---\\ \hline 
Model-5 & & & & & & & & &$\bullet$ & $\bullet$ 
&$W\supset \frac{\lambda_{HH\chi_-}^{}}{2}H_u^{}\cdot \chi_-^{}H_u
+\frac{\lambda_{HH\chi_+}^{}}{2}H_d^{}\cdot \chi_+^{}H_d$
\\ \hline 
Model-6 & & & & & &$\bullet$ &$\bullet$ & & &  
&---\\ \hline 
Model-7 &$\bullet$ & & & & &$\bullet$ &$\bullet$ & & &  
& $W\supset 
\lambda_{H_uH_dS}^{}H_u^{}\cdot H_d^{}S
+\lambda_{H_u^{\prime}H_dS}^{}H_u^{\prime}\cdot H_d^{}S
$\\
&&&&&&&&&&  
& $\phantom{W\supset} 
+\lambda_{H_u^{}H_d^{\prime}S}^{}H_u^{}\cdot H_d^{\prime}S
+\lambda_{H_u^{\prime}H_d^{\prime}S}^{}H_u^{\prime}\cdot H_d^{\prime}S
$\\ \hline 
Model-8 & & & & & &$\bullet$ &$\bullet$ &$\bullet$ & &  
& $W\supset 
\lambda_{H_uH_d\xi}^{}H_u^{}\cdot \xi H_d^{}
+\lambda_{H_u^{\prime}H_d\xi}^{}H_u^{\prime}\cdot \xi H_d^{}
$\\
&&&&&&&&&&  
& $\phantom{W\supset} 
+\lambda_{H_u^{}H_d^{\prime}\xi}^{}H_u^{}\cdot \xi H_d^{\prime}
+\lambda_{H_u^{\prime}H_d^{\prime}\xi}^{}H_u^{\prime}\cdot \xi H_d^{\prime}
$\\ \hline 
Model-9 & &$\bullet$ &$\bullet$ & & & $\bullet$&$\bullet$ & & &  
&$W\supset \lambda_{HH\Omega_-}^{}H_u^{}\cdot H_u^{\prime}\Omega_-
+\lambda_{HH\Omega_+}^{}H_d^{}\cdot H_d^{\prime}\Omega_+$
\\ \hline 
  \end{tabular}
\end{center}
%%%20110131
  \caption{Particle entries in each SSM. 
  Additional terms in the superpotential which are 
  relevant to the extra chiral superfields are also shown in 
  each model.
%%%
 }
  \label{tab:models}
 \end{table}

\section{The method}
The effective potential for the order parameter $\varphi$ 
can be written at the one-loop order as\cite{SU2U1potential}
\begin{eqnarray}
%&& \hspace{-0.8cm} 
V_{\rm eff}(\varphi)=
-\frac{\mu_0^2}{2} \varphi^2 + \frac{\lambda_0}{4} \varphi^4 %\nonumber \\
%&& \hspace{-0.2cm}
+ \sum_f 
\frac{(-1)^{2s_f} N_{C_f} N_{S_f}}{64 \pi^2}
 m_f(\varphi)^4 \left[\ln \frac{m_f(\varphi)^2}{Q^2}
-\frac{3}{2}\right], \label{eq:potential}
\end{eqnarray}
where $\mu_0^2$ and $\lambda_0$ are the bare squared mass and the coupling constant,   
$m_f(\varphi)$ is the field dependent mass of the field $f$ in the loop, and 
$N_{C_f}$ and $N_{S_f}$ are the degree of freedom of the colour and the spin with 
$s_f$ being the spin of the field $f$.
In  SSMs, the Higgs sector has a multi-Higgs structure. %,  
When the extra Higgs scalars are heavy enough, 
only the lightest Higgs boson $h$ stays at the EW scale, and behaves
as the SM-like one.
The effective potential in Eq.~(\ref{eq:potential})   
can then be applied with a good approximation.
The vacuum,  
the mass $m_h$ and the $hhh$ coupling constant  
$\lambda_{hhh}$ are determined at the one-loop order by the conditions; 
\begin{eqnarray}
\left. \frac{\partial V_{\rm eff}}{\partial \varphi}\right|_{\varphi=v} 
\hspace{-3mm}= 0, \hspace{2mm} 
\left. \frac{\partial^2 V_{\rm eff}}{\partial \varphi^2}\right|_{\varphi=v} 
\hspace{-3mm}= m_h^2, \hspace{2mm} 
\left. \frac{\partial^3 V_{\rm eff}}{\partial \varphi^3}\right|_{\varphi=v} 
\hspace{-3mm}= \lambda_{hhh}^{}, 
\label{conditions}
\end{eqnarray}
where $v$ ($\simeq 246$ GeV) is 
the vacuum expectation value for the EWSB.

\section{The Mass and the triple coupling of the lightest Higgs bosons}

\subsection{The Mass of the lightest Higgs boson}
In the MSSM, $m_h$ is evaluated by using Eq.~(\ref{conditions}) as 
\begin{eqnarray}
 m_h^2 = m_Z^2 \cos^2 2\beta
       + \delta m_{\rm loop}^2,  
\end{eqnarray}
where $\tan\beta\equiv\langle H_u^0 \rangle/\langle H_d^0 \rangle$.
The first term of the R.H.S. comes from the D-term  at the tree level. 
%,   
The quantum correction $\delta m_{\rm loop}^2$ 
has turned out to be important~\cite{mass_mssm,mssm_mh} 
to satisfy %the constraint from 
the LEP data 
($m_h > 114$ GeV)~\cite{LEP};  
\begin{eqnarray}
\delta m_{\rm loop}^2 \simeq  \frac{3m_t^4}{4\pi^2v^2}
        \ln\frac{m_{\tilde{t}_1}^2 m_{\tilde{t}_2}^2}{m_t^4} %\nonumber \\
+ \frac{3m_t^2 X_t^2 \sin^2\beta}{4\pi^2(m_{\tilde{t}_2}^2-m_{\tilde{t}_1}^2)} 
  \ln \frac{m_{\tilde{t}_2}^2}{m_{\tilde{t}_1}^2}
  +\mathcal{O}(X_t^3/m_{\tilde{t_i}}^3)
  , 
\end{eqnarray}
where $m_t$ is the top quark mass,  
$m_{\tilde{t}_{1,2}}$ are the masses of stops 
$\tilde{t}_{1}$ and $\tilde{t}_{2}$ ($m_{\tilde{t}_1}<m_{\tilde{t}_2}$),   
and 
$X_t$ is defined such that the coupling constant 
of $\tilde{t}_L$-$\tilde{t}_R$-$h$ 
is given by $X_t\sin\beta/\sqrt{2}$. 
Because the top Yukawa coupling is of order one,
$\delta m_{\text{loop}}^2$ can push up the upper bound to 
120-130GeV when stop masses are of order one TeV.
Higher order calculations for $m_h$ have been studied 
in literature\cite{highloopmh}.
%xxxxxxxxxxxx
\begin{table*}[t]
\begin{center}
  \begin{tabular}{|l|l|l|} 
   \hline\hline
 & \hspace*{0.0cm} Squared mass ($m_h^2$) of the SM-like Higgs boson $h$  &  
  \hspace*{0.0cm}Ratio of the $hhh$ coupling:  
 $\lambda_{hhh}^{\rm Model}/\lambda_{hhh}^{\rm SM}$  \\ 
\hline\hline
%SM &  \hspace{2cm} free parameter 
%& $\displaystyle \hspace{0.0cm} = 1 \frac{}{}$  \\ \hline
%2HDM & \hspace{2cm} free parameter 
%& $\displaystyle \hspace{0.0cm} \simeq
%  1 + \sum_{\phi=A,H,H^\pm} \frac{c_\phi m_{\phi}^4}{12\pi^2v^2m_h^2}
%\left(1-\frac{M^2}{m_{\phi}^2} \right)^3$ \\ \hline\hline
MSSM  &   
&  \\ 
%MSSM+$\Omega$
Model-3  & $\displaystyle \hspace{0.0cm}  \simeq  m_Z^2 \cos^22\beta +\delta m_{\rm loop}^2$  
& $\displaystyle\hspace{0.0cm} \simeq  1 + {\cal O}\left(\frac{v^2}{m_{\rm SUSY}^2}\right) $ \\ 
Model-4 &&\\
%4DSSM  
Model-6 & 
&  \\ \hline
%MSSM+$S$ 
Model-1 & $\displaystyle  \hspace{0.0cm}   \simeq  m_Z^2\cos^22\beta$
& $\displaystyle\hspace{0.0cm}  
\simeq 1  + \sum_{c=1}^2\ \frac{m_{S_c}^4}{12\pi^2v^2m_h^2}
\left(1-\frac{M_{S_c}^2}{m_{S_c}^2} \right)^3$ \\
 &
$\displaystyle\hspace{0.0cm}  + \sum_{c=1}^2\frac{m_{S_c}^2}{2}
 \left(1-\frac{M^2_{S_c}}{m_{S_c}^2}\right) \sin^22\beta
% +\delta m_{\rm loop}^2 
 +\delta \overline{m}_{\text{loop}}^2(\text{Model-1})$&\\ \hline
%MSSM+$\chi$ 
Model-5 & $\displaystyle \hspace{0.0cm}     
\simeq m_Z^2\cos^22\beta$   & \\ 
&    $\displaystyle \hspace{0.0cm}     +  \sum_{i=\pm} \sum_{c=1}^4 
  \frac{m_{T_i^c}^2}{4}\left(1-\frac{M_{T_i^c}^2}{m_{T_i^c}^2}\right) C_i 
    %+\delta m_{\rm loop}^2
 	+\delta \overline{m}_{\text{loop}}^2(\text{Model-5})$ & 
 $\displaystyle\hspace{0.0cm}  \simeq 1 + \sum_{i=\pm}\sum_{c=1}^4
\frac{m_{T_i^c}^4}{12 \pi^2 v^2 m_h^2}\left(1- \frac{M_{T_i^c}^2}{m_{T_i^c}^2}\right)^3$ \\  \hline
%4DSSM+$\Omega$
Model-9 
 & $\displaystyle  \hspace{0.0cm}   \simeq  m_Z^2
\cos^22\beta+\delta m_{\rm loop}^2$  
& $\displaystyle \hspace{0.0cm} \simeq
  1 + \sum_{i=\pm} \sum_{c=1}^2 \frac{m_{\Omega_i}^4}{12\pi^2v^2m_h^2}
\left(1-\frac{M_{\Omega_i}^2}{m_{\Omega_i}^2} \right)^3$ \\ \hline\hline
\end{tabular}
\end{center}
  \caption{ 
Formulae for $m_h$ and $\lambda_{hhh}^{\rm Model}/\lambda_{hhh}^{\rm SM}$ in various SSMs at the 
one-loop order\cite{ksy_full}. 
Masses of the component fields $S_c$, $T_i^c$, and $\Omega_i$ 
are given by $m_{S_c}$, $m_{T^c_i}$ and $m_{\Omega_i}$, while  
$M_{\phi_i^c}^2=m_{\phi_i^c}^2-\lambda_{HH\phi_i}^2 v^2 C_i/2$, where 
$\phi_i^c= S_c$, $T_i^c$, or $\Omega_i$ and 
$C_-=\sin^2\beta$ and $C_+=\cos^2\beta$ in both Model-5 and Model-9 and $C_i=1$ in the other models.
In Model-2, Model-7 and Model-8,
there are similar non-decoupling effects to Model-1.
}
 \label{tab:formulae}
 \end{table*}

\begin{figure}[t]
\begin{center}
\vspace{2cm}  \epsfig{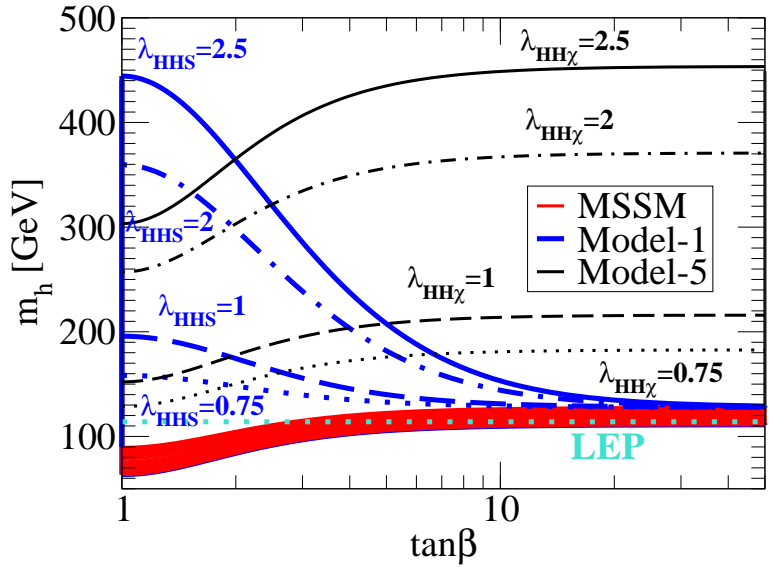}
\end{center}
 \caption{
	 The upper bounds on $m_h$ as a function of $\tan\beta$
	 for fixed values of $\lambda_{HHS}$ and $\lambda_{HH\chi_-}^{}=\lambda_{HH\chi_+}^{}\equiv \lambda_{HH\chi}$
	 in Model-1 and Model-5, respectively.
	 The red-filled region indicates the possible allowed region in the MSSM.
}
 \label{fig:tanb-dependences}
\end{figure}
In general cases with additional chiral superfields,  
the F-term can also contribute to $m_h$ at the tree level. 
For example, in Model-1 (i.e., the NMSSM), 
the term of $\lambda_{HHS} H_u \cdot H_d S$ in the superpotential 
gives the coupling of $\lambda_{HHS}^2(H_u\cdot H_d)^2$ 
in the Higgs potential, which yields   
\begin{eqnarray}
 m_h^2 \simeq m_Z^2 \cos^2 2\beta + (\lambda_{HHS}^2 v^2/2) \sin^2 2\beta 
	   +\delta \overline{m}_{\text{loop}}^2(\text{Model-1})
	   , \label{mh_nmssm}
\end{eqnarray}
where
\begin{equation}
\delta \overline{m}_{\text{loop}}^2(\text{Model-1})=
\delta m_{\rm loop}^2 +
\frac{\lambda_{HHS}^4v^2}{32\pi^2}\ln\frac{m_{S_1}^2m_{S_2}^2}{m_{\tilde{S}}^4}
\;,
\end{equation}
where 
$m_{S_1}^{}$, $m_{S_2}^{}$, and
$m_{\tilde{S}}^{}$ are
masses of 
a CP-even scalar component,
a CP-odd scalar component
and 
a fermion component of $S$, respectively.
We here assume that the mixing between component fields
are so small that each component field can be regarded as a mass
eigenstate\footnote{Through out this letter, we adopt this assumption 
also for the other models for simplicity.}.
We consider the case that the scalar and the fermion components of the 
additional fields are approximately degenerate and the logarithmic 
correction from them are negligible.
Due to the $\lambda_{HHS}^2$ term,  $m_h$ can be above 130 GeV. It is bounded from 
above by the renormalization group equation 
analysis assuming the condition of avoiding the 
Landau pole below a given value of $\Lambda$ \cite{triviality,nmssm_mh}.  
The upper bound of $m_h$ is evaluated 
to be about 140 GeV at $\tan\beta \simeq 2$ (about 450 GeV at $\tan\beta\simeq 1$) 
for $\Lambda\simeq 10^{16}$~GeV (4~TeV)\cite{ellwanger}.

Similarly, the enhancement of $m_h$ due to the F-term contribution 
can also be realized in models with an additional field $\phi$ 
where the superpotential has the gauge singlet 
operator like $H_u H_d \phi$. In Table~\ref{tab:models}, 
Model-1, Model-2, Model-5, Model-7 and Model-8 satisfy this condition, 
where $m_h$ receives the F-term contribution 
and can be significantly larger than that in the MSSM.
Approximate formulae for $m_{h}$ are given in Table~\ref{tab:formulae}.

In  Fig.~\ref{fig:tanb-dependences}, the upper bounds on $m_h$ in Model-1 and 
Model-5 are shown as a function of $\tan\beta$, and the possible allowed region in the 
MSSM is also indicated by the red-filled region.
The coupling constants $\lambda_{HH \phi}$ 
($\phi=S$ and $\chi_\pm$) are taken as 
$0 < \lambda_{HH \phi} <2.5$, whose upper limit corresponds to 
$\Lambda \gtrsim 4$ TeV in Model-1 for the case with the typical 
scale of the soft SUSY breaking to be $500$ GeV\footnote{
	The relation between the upper limit of $\lambda_{HH\phi}$ and $\Lambda$
	is not unique depending in particular on unfixed SUSY parameters.
	In addition, the running property is slightly different among the models.
	In order to examine the difference in the allowed region  among the models
	in the $m_h$-$\Delta \lambda_{hhh}^{\text{Model}}/\lambda_{hhh}^{\text{SM}}$ plane 
	with avoiding such complexity, we choose the same criterion $\lambda_{HH\phi}<2.5$ for the coupling 
	constants in each model.
%	Although it would be natural to compare models by the common cut-off scale $\Lambda$, 
	}.
%%%
The detail is shown elsewhere~\cite{ksy_full}.
%%%
In Model-1 with a fixed value of $\lambda_{HHS}^{}$, $m_h$ can be maximal for $\tan\beta = 1$, 
while in Model-5 it becomes maximal for large values of $\tan\beta$
for a fixed value of $\lambda_{HH\chi}$.
The maximal value in Model-1 becomes asymptotically the same 
as that in the MSSM in the large $\tan\beta$ limit 
up to the one-loop logarithmic contributions. %,  

\subsection{The $hhh$ coupling}
We turn to the discussion on the quantum effect on 
the $hhh$ coupling in the SSMs in the case where $h$ is
regarded as the SM like Higgs boson.
To this end, we start from the case in the non-SUSY extended Higgs sector.
It is known that in the non-SUSY two Higgs doublet model (THDM),
the $hhh$ coupling can receive large non-decoupling effects 
from the loop contribution of extra Higgs bosons, when their 
masses are generated mainly by EWSB~\cite{KKOSY,nondec}.
When $h$ is the SM like Higgs boson, physical masses
of the extra scalar bosons
are expressed by 
\begin{equation}
m_{\Phi_i}^2=M^2+\frac{\lambda_i^{} v^2}{2}\;,
\label{m2M2lamv2}
\end{equation}
where 
$\Phi_i^{}$ represents $H^0$, $H^{\pm}$ or $A^0$, and 
$M$ is the invariant mass scale which is irrelevant to the EWSB, 
and $\lambda_i^{}$ is a coupling for $\Phi_i^{\dagger} \Phi_i^{} h h$.
The physical meaning of $M$ is discussed in, for example, Ref.~\cite{KKOSY,nondec}.
The one-loop contribution to the renormalized $hhh$ coupling is calculated 
as\cite{KKOSY,nondec}
\begin{equation}
\frac{\lambda_{hhh}^{\text{THDM}}}{\lambda_{hhh}^{\text{SM}}}
\simeq 1+
\frac{1}{12\pi^2m_h^2v^2}
\left\{
m_{H^0}^4
\left(1-\frac{M^2}{m_{H^0}^2}\right)^3
+
m_{A^0}^4
\left(1-\frac{M^2}{m_{A^0}^2}\right)^3
+
2m_{H^{\pm}}^4
\left(1-\frac{M^2}{m_{H^{\pm}}^2}\right)^3
\right\}
\;.
\end{equation}
One finds that for $M^2\gg \lambda_i^{}v^2$ it becomes
\begin{equation}
\frac{\lambda_{hhh}^{\text{THDM}}}{\lambda_{hhh}^{\text{SM}}}
\simeq 
1+
\frac{v^2}{96\pi^2m_h^2}
\left(
\lambda_{H^0}^{3}
+\lambda_{A^0}^{3}
+2\lambda_{H^{\pm}}^{3}
\right)\left(\frac{v^2}{M^2}\right)
\;,
\end{equation}
which vanishes in the large $M$ limit according to 
the Appelquist-Carazzone decoupling theorem\cite{Appelquist}.
On the contrary, when the physical scalar masses 
are mainly determined by the $\lambda_i^{} v^2$ term,
the loop contribution to the $hhh$ coupling does not decouple, and 
the quartic powerlike contributions of $m_{\Phi_i^{}}^{}$ remain;
\begin{equation}
\frac{\lambda_{hhh}^{\text{THDM}}}{\lambda_{hhh}^{\text{SM}}}
\simeq 1+
\frac{1}{12\pi^2m_h^2v^2}
\left(
m_{H^0}^4
+
m_{A^0}^4
+
2m_{H^{\pm}}^4
\right)
\;.
\end{equation}
Consequently, a significant quantum effect can be realized for the $hhh$ 
coupling when $m_{\Phi_i}^2 > m_h^2$.
The size of the correction from the SM value can be of 100\% for $m_h^{}=120$GeV,
$M\simeq 0$, and $m_{H^0}^{}\simeq m_{A^0}\simeq m_{H^{\pm}}\simeq 400$GeV under the constraint from 
perturbative unitarity\cite{KKT}.
Such a large non-decoupling effect on the $hhh$ coupling 
is known to be related to the strongly first 
order EW phase transition~\cite{ewbg-thdm2}  
which is required for the EW baryogenesis\cite{ewbg-thdm}. 
\begin{figure}[t]
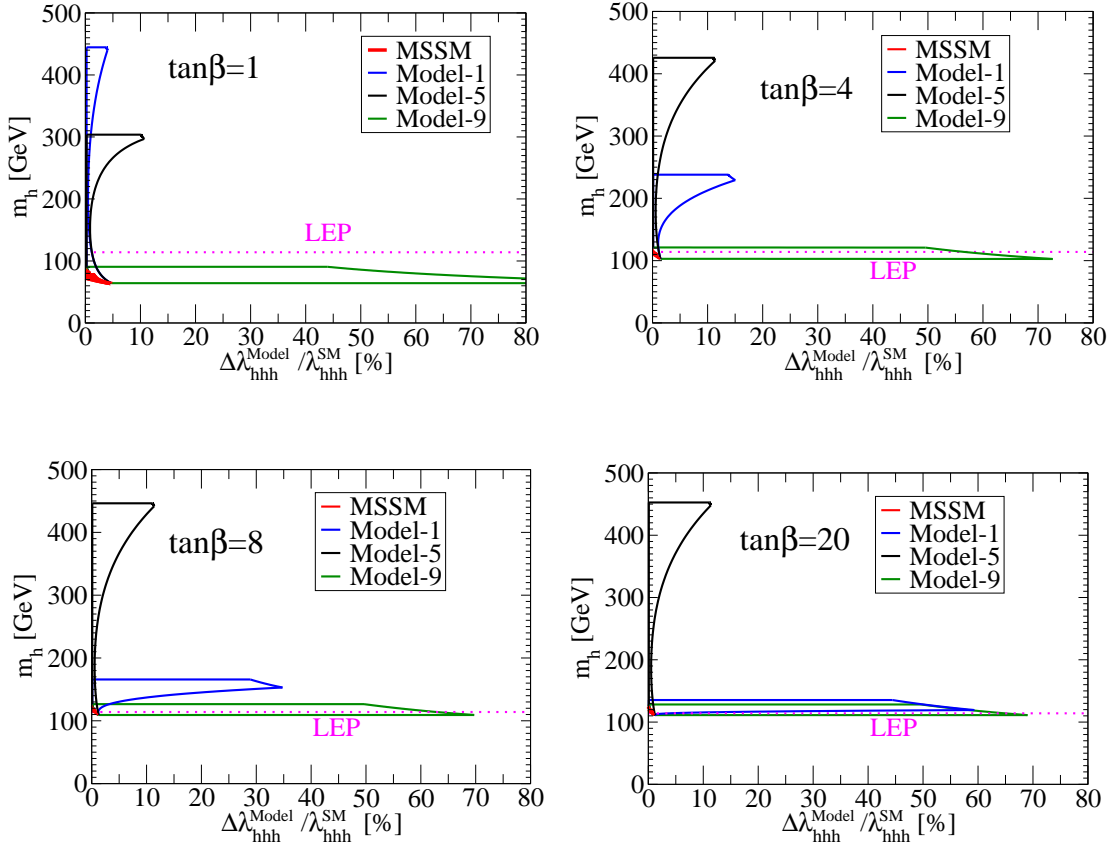

\begin{center}
%  \epsfig{file=Fig2_tanb1v4.eps,width=4cm}
%  \epsfig{file=Fig2_tanb4v4.eps,width=4cm}\\ 
%\vspace{5mm}
%  \epsfig{file=Fig2_tanb8v4.eps,width=4cm}
%  \epsfig{file=Fig2_tanb20v4.eps,width=4cm}
  \epsfig{file=Fig2_tanb1v4.eps,width=7cm} \hspace{2mm}
  \epsfig{file=Fig2_tanb4v4.eps,width=7cm}\\ 
\vspace{1cm}
  \epsfig{file=Fig2_tanb8v4.eps,width=7cm}\hspace{2mm}
  \epsfig{file=Fig2_tanb20v4.eps,width=7cm}
\end{center}
%%%20110131
 \caption{
Possible allowed regions in the $m_h$-$(\Delta \lambda_{hhh}/\lambda_{hhh})$ plane  
in the MSSM, Model-1, Model-5 and Model-9 for each $\tan\beta$ value.
We scan the parameter space as
$0<\lambda_{HH\phi}<2.5$, 
$0.5\,\text{TeV}<m_{\tilde{t}_{1,2}}^{}<1.5\,\text{TeV}$,
and
$0.5\,\text{TeV}<m_{\phi_i^c}^{}$
for each model.
}
%%%20110131
 \label{fig:mh-hhh_tanb}
\end{figure}
%
% 
%
%
%
%\section{Analysis and results}

Let us discuss the $hhh$ coupling in the SSMs listed in Table~\ref{tab:models}. 
In the MSSM, it is evaluated at the one-loop level as 
\begin{eqnarray}
\lambda^{\rm{MSSM}}_{hhh}\simeq
\frac{3m_h^2}{v} \left[    1  -  \frac{m_t^4}{\pi^2v^2m_h^2}   
    \left\{ 
 1  -   \frac{m_t^2 (m_{\tilde{t}_1}^2+m_{\tilde{t}_2}^2)}{2 m_{\tilde{t}_1}^2 m_{\tilde{t}_2}^2} 
 + \frac{3X_t^2v^2\sin^2\beta}{4m_{\tilde{t}_1}^2m_{\tilde{t}_2}^2}
    \right\}
\right], 
\end{eqnarray}
up to the stop mixing~\cite{ewbg-thdm2}. 
This result coincides with that in the SM   
in the decoupling limit $m_{\tilde{t}_{1,2}} \to \infty$~\cite{KKOSY,nondec}. % ,   
On the contrary, in a class of the SSMs; Model-1, Model-5 and Model-9,
the non-decoupling effect can appear in the $hhh$ 
coupling, 
similarly to the case of the THDM.
Approximate formulae of the $\lambda_{hhh}^{\rm Model}/\lambda_{hhh}^{\rm SM}$ 
are given in Table~\ref{tab:formulae}.
In the following, we explain the results in each model in order.

In Model-1 a scalar boson $S_1$ and a pseudo-scalar boson $S_2$ from 
extra isospin singlet field $S$ are running in the one-loop diagrams
of the $hhh$ coupling.
Their physical masses $m_{S_c}^2$ $(c=1,2)$ are given by 
\begin{equation}
m_{S_c}^2=M_{S_c}^2+\frac{1}{2}\lambda_{HHS}^2v^2\;,
\end{equation}
where $M_{S_c}$ represent the invariant mass parameters
and $\lambda_{HHS}$ is a coupling constant
of the $H_u\cdot H_dS$ in the superpotential.
The loop effect decouples in the large $M_{S_c}^2$ limit
in the same way as the THDM.
The non-decoupling property appears when 
$\lambda_{HHS}^2v^2\gtrsim M_{S_c}^2$.
However, differently from the THDM, $M_{S_c}^2$ may not be too small,
because this is directly related to the mass of the singlino.
When $M_{S_c}$ is taken to be around 500 GeV, the one-loop contribution
still turns out to be important. 
For example if $M_{S_c}\simeq 500$ GeV, $\lambda_{HHS}\simeq 2.5$ and $\tan\beta\simeq 20$ ($m_{S_c}\simeq 660$ GeV),
the correction to the SM prediction can be as large as 40\%.
Here we take $m_{\tilde{S}}\simeq m_{S}$.
The quantum correction to the $hhh$ coupling in Model-1 strongly 
depends on $\tan\beta$ because the lightest Higgs boson mass $m_h$ 
depends on $\tan\beta$ (see Fig.~\ref{fig:mh-hhh_scaned}).

As is seen in Table~\ref{tab:formulae}, similar effect can also be
realized in Model-5 and Model-9. 
In Model-5 there are two types of the triplet fields ($T_-^{}$ and $T_+^{}$), 
each of which 
gives six degrees of freedom, namely two neutral scalar bosons,
a pair of singly charged Higgs bosons, and a pair of doubly 
charged Higgs bosons. However, only neutral and singly charged 
degrees of freedom contribute to the loop effect of the $hhh$ coupling,
because a term like $hhT^{++}T^{--}$ cannot exist in the Higgs potential.
In this formula, for each $i$ ($=-$ or $+$),
$T_{i}^1$ is the CP-even scalar boson, 
$T_{i}^2$ is the pseudo-scalar boson, and $(T_{i}^3\pm i T_{i}^4)/\sqrt{2}$ 
represents
a pair of the singly charged Higgs bosons.
Their physical masses $m_{T_i^c}$ are 
\begin{equation}
m_{T_i^c}^2=M_{T_i^c}^2+\frac{1}{2}\lambda_{HH\chi_i}^2v^2C_i\;,
\end{equation}
where $M_{T_i^c}$ are the invariant mass parameter, 
and $C_-^{}=\sin^2\beta$ and $C_+^{}=\cos^2\beta$. 
Here $m_{T_i^3}=m_{T_i^4}$ ($\equiv m_{T_i^{\pm}}$) 
and $M_{T_i^3}=M_{T_i^4}$ are assumed.
One might think that similarly to the THDM and Model-1, 
the correction to the $hhh$ coupling can be significant,
when $m_{T_i^c}$ is mainly from the $\lambda_{HH\chi_i}v^2$ term.
However, this is not the case because 
$m_h$ becomes also large by the same mechanism of enhancement as 
the $hhh$ coupling so that the net correction cannot be very significant.

Next we discuss the case of Model-9 where 
there are two extra doublets and two singly charged singlets.
Notice that the extra doublets do not contribute to the one loop 
correction to the $hhh$ coupling
because there is no corresponding F-term, so that
only the one-loop effect of charged singlets can be important.
Two singly charged scalar bosons ($\Omega_i^{}$ with $i=-$ or $+$)
are running in the one-loop diagram of the $hhh$ coupling.
Their physical masses $m_{\Omega_{i}^{}}$ are given by 
\begin{equation}
m_{\Omega_{i}^{}}^2=M_{\Omega_i^{}}^2+\frac{1}{2}\lambda_{HH\Omega_i}^2v^2C_i\;,
\end{equation}
where $M_{\Omega_i^{}}$ are the invariant mass parameters.
Differently from Model-5, we can expect large deviation in the $hhh$ 
coupling from the SM prediction in this model, because $m_h$ does not
get a significant enhancement from the F-term contribution.
We have similar large effect to that in Model-1 and THDM,
when $m_{\Omega_i^{}}$ is mainly from $\lambda_{HH\Omega_i^{}}^2v^2$.

In summary,
a non-decoupling quartic power-like scalar mass effect occurs in the $hhh$ coupling 
when additional scalar field $\phi$ (such as 
$S_{1,2}$, $H_{u,d}'$, $T_{\pm}^{0,+}$, etc.) receives 
its mass mainly from $v$ via the operators 
 $H_i H_j \phi \phi$,
which are generated from $H_i H_j \phi$ in the superpotential, 
where $H_i$ ($i=u,d$) are the Higgs doublets.  
This effect appears in Model-1, Model-2, Model-5, Model-7, Model-8 
and Model-9, while there is no such contribution in the other models. 
Even in models with this effect,
the deviation from the SM prediction in the $hhh$ coupling cannot be
significant,
when $m_h$ is also enhanced by the same F-term contribution. % simultaneously. 
%%%20110131
For example, in Model-1  with a small $\tan\beta$ ($\sim 1$),
large $\lambda_{HHS}^{}$ gives a significant contribution to the 
numerator. However at the same time, the $m_h^{}$ in the denominator 
gets the large contribution as seen in Eq.~(\ref{mh_nmssm}).
Consequently, 
enhancement by the non-decoupling effect of particle $S$ are weakened,
and the deviation in the $hhh$ coupling is not very important.
Model-2, Model-5, Model-7 and Model-8 also correspond to this case.
%%%20110131
%
On the other hand, in Model-9 for example, 
the F-term of the operator $H_u \cdot H_u' \Omega_-$ or $H_d \cdot H_d' \Omega_+$ 
cannot contribute to $m_h$ but
gives a quartic power contribution of $m_{\Omega_{+/-}}$ in
the one-loop corrected $hhh$ coupling, 
so that the deviation in the $hhh$ coupling from the SM value can be significant. 

\subsection{The correlation between $m_h^{}$ and the $hhh$ coupling}
%%%20110131
We scan the parameter space in each model to find allowed regions in the
$m_h$-$(\Delta \lambda_{hhh}^{\rm Model}/\lambda_{hhh}^{\rm SM})$ plane 
under the assumption of $\lambda_{HH\phi}<2.5$ at the EW scale,
where $\Delta \lambda_{hhh}^{\rm Model}=\lambda_{hhh}^{\rm Model}-\lambda_{hhh}^{\rm SM}$.
In Fig.~\ref{fig:mh-hhh_tanb}, we show the possible allowed region 
%in the $m_h$-$(\Delta \lambda_{hhh}^{\rm Model}/\lambda_{hhh}^{\rm SM})$ plane 
for several value of $\tan\beta=1$, $4$, $8$ and $20$.
%where $\Delta \lambda_{hhh}^{\rm Model}=\lambda_{hhh}^{\rm Model}-\lambda_{hhh}^{\rm SM}$.
The coupling constants $\lambda_{HH\phi}$ % of $H_i H_j\phi$ 
($\phi=S$, $\chi_\pm$ and $\Omega_\pm$) are taken to be less than 2.5 as 
in Fig.~\ref{fig:tanb-dependences}.
The stop masses are scanned as $0.5\,\text{TeV}\leq m_{\tilde{t}_{1,2}}^{}\leq 1.5\,\text{TeV}$.
We also scan the physical masses of the extra scalar bosons as $0.5\,\text{TeV}\leq m_{\phi}^{}$.
The mass of fermion component is taken as same as the mass of the scalar component
for each extra field.
We note that the parameters are scanned such that the additional contributions to the 
rho parameter are negligible\footnote{For example, parameters in the stop-sbottom sector are 
taken to keep the rho parameter constraint satisfied.}.
%%%20110131
The region in the MSSM is indicated as the red-filled one. 
The possible allowed region in Model-1 depends largely on $\tan\beta$: 
for smaller (larger) $\tan\beta$, $m_h$ can be higher (lower) and  
$\Delta\lambda_{hhh}^{\rm Model\mbox{-}1}/\lambda_{hhh}^{\rm SM}$ is smaller (larger).
Model-5 is relatively insensitive to the value 
of $\tan\beta$: $m_h$ can always be larger than 
about 300 GeV while $\Delta\lambda_{hhh}^{\rm Model\mbox{-}5}/\lambda_{hhh}^{\rm SM}$ 
remains less than about 10 \%.
On the other hand, in Model-9, although the possible value of $m_h$ 
is similar to that in the MSSM, % because of no F-term contribution, 
the deviation in the $hhh$ coupling can be very large: 
i.e.,  $\Delta\lambda_{hhh}^{\rm Model\mbox{-}9}/\lambda_{hhh}^{\rm SM} \sim 30-60$ \%\footnote{
The definition of $\tan\beta$ in models with four Higgs doublets 
is that %rather different from those with two Higgs doublets; i.e., 
$\tan\beta = \sqrt{\langle H_u^0 \rangle^2+\langle H_u'^0 \rangle^2}/
\sqrt{\langle H_d^0 \rangle^2+\langle H_d'^0 \rangle^2}$.}. 
When we consider the higher value of $\Lambda$, which corresponds to the smaller upper bound on 
$\lambda_{HH\phi}^{}$, the possible allowed region becomes the smaller.

\begin{figure}[t]
\begin{center}
\vspace*{0.2cm}
  \epsfig{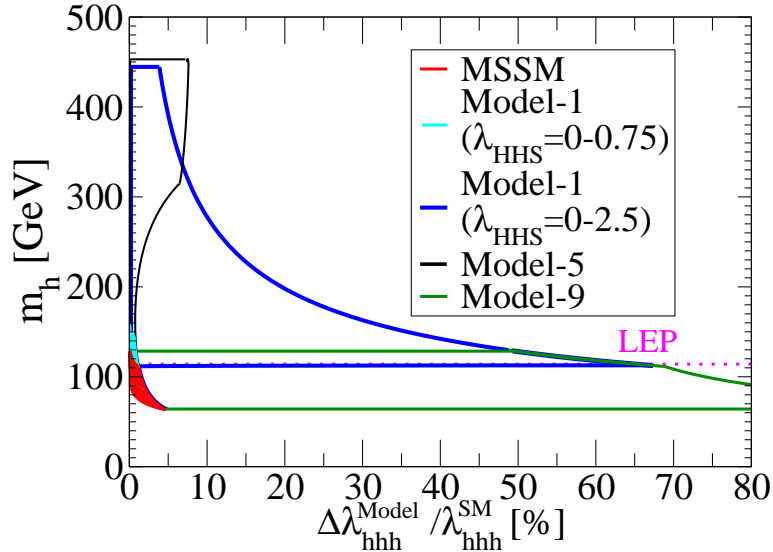}
\end{center}
 \caption{
Possible allowed regions in the $m_h$-$(\Delta \lambda_{hhh}/\lambda_{hhh})$ plane  
in the MSSM, Model-1, Model-5 and Model-9 with scanned $\tan\beta$.}
 \label{fig:mh-hhh_scaned}
\end{figure}

In Fig.~\ref{fig:mh-hhh_scaned}, 
possible allowed regions with scanned $\tan\beta$ are shown in the 
$m_h$-$(\Delta \lambda_{hhh}^{\rm Model}/\lambda_{hhh}^{\rm SM})$ plane 
in Model-1, Model-5 and Model-9 as well as the MSSM. 
The maximal values of $\lambda_{HH\phi}^{}$ in Model-1, Model-5 
and Model-9 are taken to be the same as those in Fig.~\ref{fig:mh-hhh_tanb}. 
The region in the MSSM (Model-1 with $0\lesssim \lambda_{HHS} \lesssim 0.75$, which corresponds 
to $\Lambda \simeq 10^{16}$ GeV~\cite{ellwanger}) 
is indicated as the red-filled (cyan-filled) one. 
The possible allowed regions 
are different among the models so that 
the information of $m_h$ and $\Delta \lambda_{hhh}^{\rm Model}$ can be used to
classify the SSMs.

\section{Discussion and Conclusion}

We have studied $m_h$ and
the $hhh$ coupling
at the one-loop level
in various SSMs,
where $h$ is the lightest SM like Higgs boson.
In a class of SSMs, the mass of the lightest Higgs boson,
which is lower than 120-130~GeV in the MSSM, 
can be much higher due to the F-term contribution.
Such an enhancement appears in the models with 
the extra neutral singlet or the triplet superfield.
The upper bounds on $m_h$ are determined by the size of the coupling constants 
of the F-term, which can be constrained by the renormalization 
group equation analysis with an imposed cut-off scale $\Lambda$.
Consequently, $m_h$ can be higher than 300-400~GeV when $\Lambda$
is at the TeV scale in Model-1 with small $\tan\beta$ values and 
Model-5.

Although the one-loop correction to the $hhh$ coupling 
due to the extra scalar components vanishes in the decoupling limit,
it can be significant in particular SSMs such as Model-1 with large 
$\tan\beta$ and Model-9, when 
$\lambda_{HH\phi_i}^2v^2 \sim M_{\phi_i}^2$
where $M_{\phi_i}^{}$ is the invariant mass parameter of the 
extra field $\phi_i$ in the loop.
In such a case, quartic powerlike mass 
contributions can appear as non-decoupling effects, and
the correction can be larger than several times ten percent under 
the constraint from parturbativity.
%%%20110131
In this letter the analysis has been restricted in the effective potential 
method, where all the external momenta are set on zero.
In the actual measurement of the $hhh$ coupling, one might think that 
the momentum dependences would be important.
For example, at the LHC the $hhh$ coupling may be 
measured by $W$ fusion process $W^{+*}W^{-*}\to h^*\to hh$\cite{hhh_lhc}, 
where the measured $hhh$ coupling is a function of $\sqrt{\hat{s}}$
the energy of the elementary process.
At the ILC and its $\gamma\gamma$ option,
the processes $e^+e^- \to Z^\ast \to Z h^\ast \to Zhh$\cite{hhh_lc}
and $\gamma\gamma \to h^\ast \to hh$\cite{gammagamma} can be used.
The energy dependence of the $hhh$ coupling has been discussed 
in Ref.~\cite{KKOSY}: see Fig.~3 in it.
It is shown that unless $\sqrt{\hat{s}}\gg 2m_h$, the energy dependence is small
in the bosonic loop contributions to the $hhh$ coupling~\footnote{
    Notice that the process $\gamma\gamma \to hh$ is one-loop induced,
    so that the correction to the hhh coupling corresponds to the two loop effect.
    For this process, the result with the energy dependence in the $hhh$ coupling
    and that without the energy dependence by using the effective $hhh$ coupling
    are given in Ref.\cite{Asakawa:2008se}.
    }.
In addition, the non-decoupling effect in the wave function renormalization 
is at most quadratic instead of quartic in mass.
Hence the wave function correction to the $hhh$ coupling is known to be
as large as of order one percent and it is negligible\cite{nondec}.
Consequently the calculation by 
using the effective potential gives a good approximation for our analysis.
%%%20110131
%

In conclusion, even when only $h$ is observed in future, 
precision measurements of $m_h$ and the $hhh$ coupling 
can help discriminate the SSMs.
Such discrimination can be improved if extra information 
for $\tan\beta$ can be used from, for example, future 
flavour experiments such as those for $B_s\to \mu\mu$, 
$B\to \tau\nu$ and so on.

In any case, the $hhh$ coupling is required 
to be measured with ${\cal O}(10)$ \% accuracy, 
which may be expected at future colliders such as the LHC upgrade, 
the ILC and the CLIC~\cite{hhh_lc}. 

\begin{acknowledgments}
We would like to thank Yasuhiro Okada for useful discussions. 
This work was supported in part by Grant-in-Aid for Scientific Research, 
Japan Society for the Promotion of Science (JSPS), 
Nos. 22244031 and 19540277~(S.K.), 
and No. 22011007~(T.S.).
The work of K.Y. was supported by JSPS Fellow (DC2). 

\end{acknowledgments}

%\vspace*{-4mm}

\end{document}